\documentclass[conference]{IEEEtran}
\IEEEoverridecommandlockouts

\usepackage{cite}
\usepackage{amsmath,amssymb,amsfonts}
\usepackage{algorithmic}
\usepackage{graphicx}
\usepackage{textcomp}
\usepackage{xcolor}
\usepackage{booktabs}
\usepackage{multirow}
\usepackage{colortbl}
\usepackage{makecell}
\usepackage{url}
\usepackage{enumitem}
\usepackage[hidelinks]{hyperref}
\definecolor{mydarkred}{rgb}{0.6,0,0}

\def\BibTeX{{\rm B\kern-.05em{\sc i\kern-.025em b}\kern-.08em
    T\kern-.1667em\lower.7ex\hbox{E}\kern-.125emX}}

\newcommand{\MYhref}[3][mydarkred]{\href{#2}{\color{#1}{#3}}}
    
\begin{document}

\title{Beyond Static LLM Policies: Imitation-Enhanced Reinforcement Learning for Recommendation
\thanks{\textsuperscript{*} indicates equal contribution.}}

\author{
Yi Zhang\textsuperscript{1,2,*}, 
Lili Xie\textsuperscript{*}, 
Ruihong Qiu\textsuperscript{1}, 
Jiajun Liu\textsuperscript{2,1}, 
Sen Wang\textsuperscript{1} \\
\IEEEauthorblockA{\textsuperscript{1}School of Electrical Engineering and Computer Science, The University of Queensland, Brisbane, Australia} 
\IEEEauthorblockA{\textsuperscript{2}Data61, CSIRO, Brisbane, Australia} 
\IEEEauthorblockA{uqyzha91@uq.edu.au, xaiverlilix@gmail.com, r.qiu@uq.edu.au, jiajun.liu@csiro.au, sen.wang@uq.edu.au}}

\maketitle

\begin{abstract}
Recommender systems (RecSys) have become critical tools for enhancing user engagement by delivering personalized content across diverse digital platforms. Recent advancements in large language models (LLMs) demonstrate significant potential for improving RecSys, primarily due to their exceptional generalization capabilities and sophisticated contextual understanding, which facilitate the generation of flexible and interpretable recommendations. However, the direct deployment of LLMs as primary recommendation policies presents notable challenges, including persistent latency issues stemming from frequent API calls and inherent model limitations such as hallucinations and biases. To address these issues, this paper proposes a novel offline reinforcement learning (RL) framework that leverages imitation learning from LLM-generated trajectories. Specifically, inverse reinforcement learning is employed to extract robust reward models from LLM demonstrations. This approach negates the need for LLM fine-tuning, thereby substantially reducing computational overhead. Simultaneously, the RL policy is guided by the cumulative rewards derived from these demonstrations, effectively transferring the semantic insights captured by the LLM. Comprehensive experiments conducted on two benchmark datasets validate the effectiveness of the proposed method, demonstrating superior performance when compared against state-of-the-art RL-based and in-context learning baselines. The code can be found at \MYhref{https://github.com/ArronDZhang/IL-Rec}{https://github.com/ArronDZhang/IL-Rec}.
\end{abstract}

\begin{IEEEkeywords}
Inverse Reinforcement Learning, Recommender Systems, Large Language Models
\end{IEEEkeywords}

\section{Introduction}
Recommender systems (RecSys) have become an essential component of digital platforms, reshaping user experiences in domains such as e-commerce, social media, and online entertainment~\cite{qiu2019rethinking,qiu2020gag,qiu2021memory,qiu2022contrastive,li2021discovering}. By analyzing user behaviors and preferences, RecSys can deliver highly personalized content, thereby enhancing user engagement and retention. Traditional RecSys, ranging from collaborative filtering\cite{marlin2009yahoo} to deep learning-based approaches~\cite{kang2018sasrec,sinha2022s4rl}, have significantly improved personalization capabilities. However, these methods often struggle with adapting to rapidly evolving user interests and optimizing recommendations for long-term user satisfaction, thereby necessitating more advanced solutions~\cite {chen2023opportunities}.

Recent advancements in large language models (LLMs)~\cite{touvron2023llama2,guo2025deepseek,achiam2023gpt4} have opened new opportunities for improving RecSys with superior generalization and contextual understanding abilities~\cite{wu2024llm4rec_survey}. LLM-based recommenders can process diverse data modalities, including text, images, and structured data, to generate more informed and flexible recommendations. Additionally, LLMs enable interpretability by generating natural language explanations for recommendations with transparency and user-friendliness. Despite these advantages, integrating LLMs into RecSys still poses significant challenges, such as high computational costs, reliance on external API calls, and the difficulty of adapting static LLM-generated policies to dynamic user preferences~\cite{yue2023llamarec}.

\begin{figure}[!t]
\centering
\includegraphics[trim=0cm 0cm 0cm 0cm, clip, width=1\linewidth]{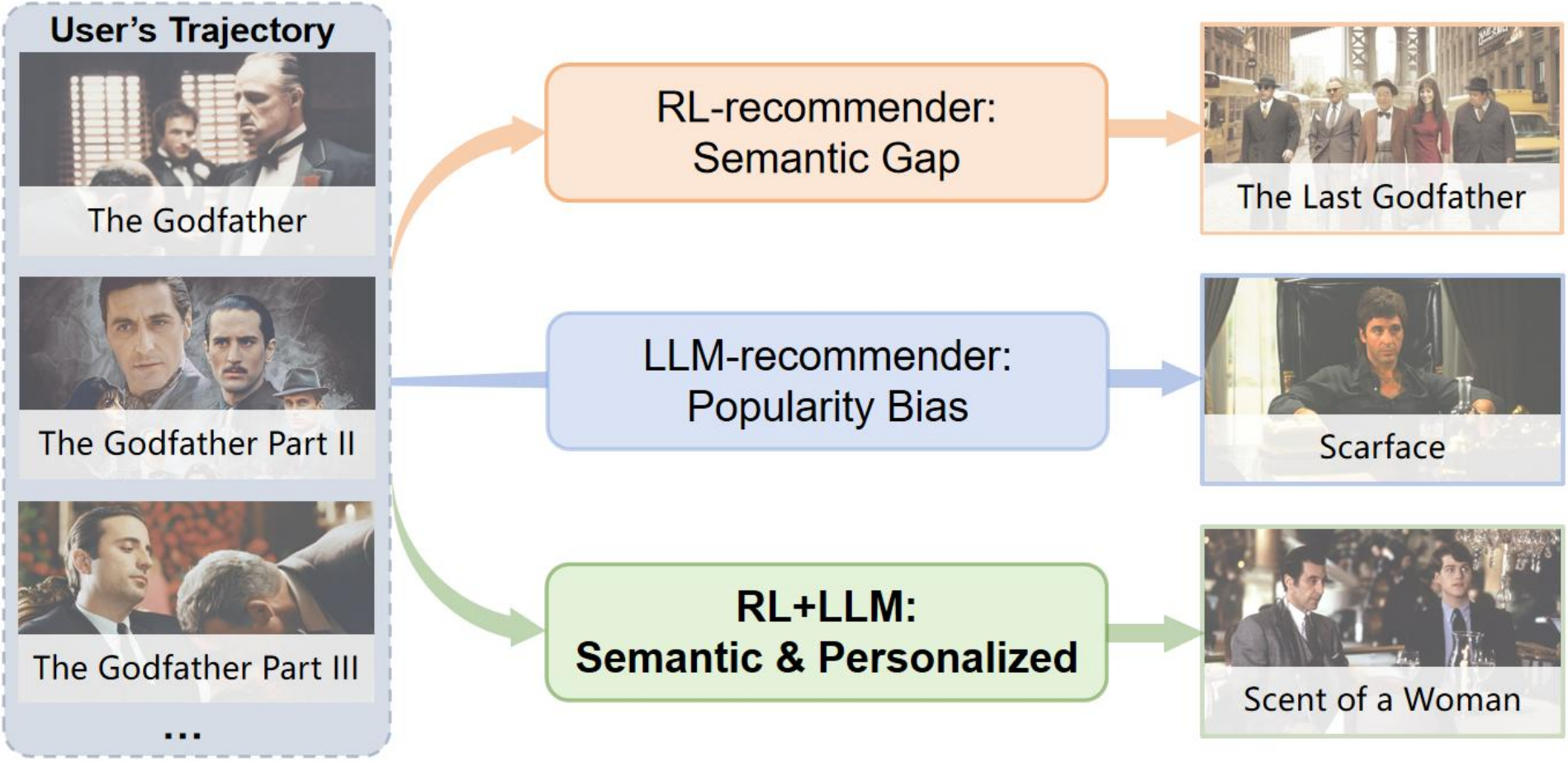}
\caption{Comparison of three recommendation strategies. RL4RS models often struggle to interpret user intent from textual input, resulting in less accurate suggestions. Static LLM policies tend to favor prevalent items of the same type, which may reduce recommendation diversity. The integration of RL into LLM recommendation exploits language understanding with adaptive and user-specific decision-making.}
\label{fig:intro}
\end{figure}

While most existing LLM-based methods utilize LLMs directly as recommendation policies~\cite{shi2024billp,yao2023react}, two primary challenges arise in practice: (1) Online recommendation demands low-latency responses~\cite{cui2024distillation,wu2024llm4rec_survey}, yet relying solely on LLMs necessitates frequent calling of the API, resulting in \textbf{consistently} high latency; (2) outputs are prone to spurious associations and reinforcement of recurrent patterns, undermining their reliability as standalone recommendation policies~\cite{huang2024hallu_survey,gao2024sprec}. Conversely, reinforcement learning (RL)‐based sequential decision approaches have been adopted to optimize long‐term engagement, but they often rely on coarse feedback signals and fail to capture nuanced semantic dependencies in user histories. As illustrated in the movie recommendation example in Fig.~\ref{fig:intro}, when a user’s interactions center on \textit{The Godfather Trilogy}, standard reinforcement learning for recommender system (RL4RS) methods may replicate substring matches (e.g., “Godfather”) without recognizing deeper context such as recurring cast members. Recent efforts~\cite{gao2024sprec,cai2025puma,shi2024billp} seek to mitigate these issues by integrating the generative strengths of LLMs with the sequential decision-making capabilities of reinforcement learning. Nevertheless, methods involving LLM fine-tuning~\cite{gao2024sprec,cai2025puma} incur substantial computational overhead and are time-intensive. Meanwhile, approaches that incorporate historical feedback into subsequent trajectories through prompting~\cite{shi2024billp} partially mimic gradient-based learning but still suffer from persistent latency and produce \textbf{sub-optimal} recommendation policies.

To address these challenges, an RL4RS approach, \textbf{IL-Rec}: Imitation‐Enhanced Reinforcement Learning for Recommendation, is proposed in which LLM-generated demonstrations are imitated to eliminate continual online LLM invocation and mitigate inherent hallucinations and biases. Specifically, in-context LLM trajectories are employed as demonstrations, removing the need for LLM fine-tuning and reducing computational overhead. Reward models corresponding to the LLM policy are extracted using inverse reinforcement learning. Simultaneously, an RL policy is trained, guided by cumulative rewards of the LLM-generated trajectories, which are used as importance weights. In this way, the semantic information captured by the LLM is leveraged, the RL training process is accelerated, and the influence of hallucinations and biases is alleviated. The proposed method is validated on two benchmark datasets and shown to outperform state-of-the-art RL4RS and in-context learning baselines.
The key contributions of this paper are summarized as follows:
\begin{itemize}
\item Limitations of LLM-based in-context learning for RecSys are identified, and reinforcement learning is leveraged to establish a low-latency, semantically informed recommendation policy framework.
\item An inverse reinforcement learning method is utilized to extract reward models from LLM demonstrations, accelerating RL policy training and enabling superior performance beyond the original LLM demonstrations.
\item Extensive experiments on two benchmark datasets are conducted, demonstrating the proposed method consistently outperforms state-of-the-art RL4RS and strong static LLM baselines.
\end{itemize}

\section{Related Work}
\subsection{Reinforcement Learning for RecSys}
Owing to the sequential decision nature of RecSys, a recommendation task can be formulated as a Markov Decision Process (MDP). In the formulation of RL, some works try to figure out a practical way to encode relevant information into the state representation~\cite{kang2018sasrec,huang2022state_repr}. Some focus on adapting existing RL methods to tackle special challenges in RecSys, such as huge action space~\cite{chen2019top} and data sparsity~\cite{chen2021user}. As training RL policies online is expensive, offline RL appears to be a feasible solution~\cite{levine2020offline,zhang2025darlr}. CIRS~\cite{gao2023cirs} introduces causal inference to capture the user preference and mitigate the filter bubble problem. DORL~\cite{gao2023dorl} proposes an effective entropy penalty to encourage offline exploration during recommendation policy learning, alleviating the Mathew effects. ROLeR~\cite{yi2024roler} identifies the inaccuracy of offline reward function estimation, exploiting a non-parametric reward shaping method to refine the world models. Though these methods show promising performance, the abundant text information in offline data is seldom utilized.

\subsection{Large Language Models for RecSys}
Diving into the practical scenarios of RecSys, such as movie platforms~\cite{harper2015movielens}, short-video platforms~\cite{gao2022kuairec,gao2022kuairand}, and music platforms~\cite{brost2019music_dataset}, rich text can be found, including the genres, introduction of the movies or music. This information provides the potential to deal with the typical challenges in RecSys, such as cold-start problems~\cite {wu2024llm4rec_survey,sanner2023llm_cold}. LLMs can process vast amounts of textual data~\cite{guo2025deepseek,achiam2023gpt4,touvron2023llama2}, capturing nuanced user preferences and providing contextually relevant suggestions. PrefRec~\cite{wanqi2023prefrec} proposes to learn reward functions from offline trajectories with RLHF~\cite{stiennon2020rlhf}. RLMRec~\cite{ren2024llm_repr} utilizes LLMs to learn informative representations to facilitate recommendations. FaiRLLM~\cite{zhang2023gpt_fair} focuses on the fairness issue in LLM for RecSys. Though RecSys benefits from embracing LLMs, challenges are also introduced, including high computational costs and the need for effective fine-tuning to align with specific recommendation tasks.

\subsection{Imitation Learning for RecSys}

While reinforcement learning (RL) has been widely adopted in sequential recommendation to optimize long-term user engagement, it often suffers from reward misalignment, where the designed reward signals fail to reflect users’ true preferences. To address this, imitation learning (IL) has emerged as a promising alternative by learning user policies directly from expert behavior. Behavioral Cloning (BC), a simple yet effective IL method, has been applied in recommender systems to mimic user actions from historical trajectories. For example, Stratified Expert Cloning (SEC)\cite{lin2025stratified} introduces expert stratification to enhance long-term retention. Advanced IL methods, such as Inverse Reinforcement Learning
(IRL)~\cite{ng2000algorithms}, which infers reward functions behind expert behaviors and Adversarial Imitation Learning (AIL)~\cite{ho2016generative}, which leverages generative adversarial networks to directly learn the policy and reward function simultaneously, extends IL's abilities. InvRec\cite{chen2021generative} combines IRL and AIL via a Generative Adversarial Imitation Learning (GAIL)-based framework, while SHIL~\cite{wang2022hierarchical} decomposes complex tasks using hierarchical imitation.

Recent studies adopt imitation as a knowledge transfer paradigm to implicitly capture user intent in offline settings. DLLM2Rec~\cite{cui2024distillation} distills knowledge from LLMs to lightweight recommenders, and LEA~\cite{wang2024rl4rs_env} treats LLMs as simulators to support RL training. Unlike prior work relying on optimal demonstrations, the proposed method leverages LLMs to generate diverse, sub-optimal trajectories and applies GAIL to learn robust policies from imperfect data, reducing dependence on expert supervision and lowering inference costs.

\section{Preliminaries}
\subsection{Model-based Offline RL}
This part details the concepts of model-based offline reinforcement learning (RL). An RL task can be formulated as a Markov Decision Process (MDP) and described by a 5-element tuple $\mathbf{G} = <\mathbf{S}, \mathbf{A}, T, r, \gamma>$~\cite{sutton2018reinforcement}. Within the MDP, $\mathbf{S}$ stands for the state space, which is the set of all possible states. Similarly, $\mathbf{A}$ is the action space defining all possible actions. $\mathbf{s} \in \mathbf{S}$ or $\mathbf{a} \in \mathbf{A}$ denote one state or action of the current MDP. $T$ is the transition function describing the dynamics of the environment. In deterministic settings such as in RecSys, $\mathbf{s_{t+1}} = T(\mathbf{s_t}, \mathbf{a_t})$ offers the next state given current state and action. $r$ is the reward function, $r(\mathbf{s_t, a_t})$, which stands for the instant reward after conducting action $\mathbf{a_t}$ in state $\mathbf{s_t}$. $\gamma$ is the discount factor balancing the current reward with future rewards. The goal of a MDP is to learn a policy $\pi$ which can maximize the cumulative reward, \textit{i.e.}, $G_t = \sum_{\mathbf{s}=\mathbf{s_0}}^{s_T} \gamma^t r(\mathbf{s_t}, \mathbf{a_t})$, where $\mathbf{s_0}$ and $\mathbf{s_T}$ are the initial and terminal states. The learning of policies usually requires quantities of interactions between the policies and the environments. In some scenarios like RecSys, online interactions are expensive while offline logs are easily available.

Leveraging offline data to mitigate the reliance on costly online interactions forms the foundation of offline RL, where the data consists of interactions, \textit{i.e.} $D = \{(\mathbf{s_t, a_t, r_t, s_{t+1}})\}^T_{t=0}$,  between the environment and behavior policies. 
Model-based offline RL methods simulate the online environments with offline data, \textit{i.e.}, $D$, by explicitly modelling the transition functions, \textit{i.e.}, $\hat{T}$, or reward functions,  \textit{i.e.}, $\hat{r}$. These simulated environments are called world models. The policies are learned by interacting with the world models as online policy training.

\subsection{Inverse RL}
Adversarial imitation learning (AIL)~\cite{ho2016generative} is employed in this work, where a discriminator, \textit{i.e.,} $D$, is utilized to differentiate the learning policy from the expert policy. And the objective of the learning policy is to "deceive" the discriminator that its trajectories are the expert demonstrations. Thus, the reward of the learning policy comes from the incorrectness of the discriminator, and that of the discriminator comes from the classification correctness. The objective of the two agents is:
\begin{equation}
\begin{split}
    \min_{\theta}\max_{\psi}& \mathbb{E}_{(s,a)\sim\rho_{\pi_\theta}}[logD_{\psi}(s,a)] \\ & + \mathbb{E}_{(s,a)\sim\rho_{\pi_E}}[log(1-D_{\psi}(s,a))],
\end{split}
\label{eq:obj}
\end{equation}
where $\theta$ and $\psi$ denote the parameters of the learning policy and the discriminator, respectively. $\rho$ denotes the corresponding distribution of the policy.

\subsection{Problem Formulation}
As the recommendation process can be regarded as a sequential decision task, it can be formulated by an MDP and tackled with RL methods. In RecSys, an action $\mathbf{a} \in \mathbf{A}$ represents recommending a specific item, and a state $\mathbf{s} \in \mathbf{S}$ details the conditions before recommending this item. The modelling of states in RL for RecSys is widely studied~\cite{huang2022state_repr}. In this paper, a common way is adopted which takes the recently interacted items and side information of users, such as gender and age into consideration. The transition function of RecSys is usually handled by a sequence model, for instance, GRU~\cite{kang2018sasrec} and Transformer~\cite{vaswani2017attention}, which encodes the item information into $\mathbf{s_t}$ to obtain $\mathbf{s_{t+1}}$. The reward functions of RecSys are often modelled by the immediate feedback after recommending an item to a user. They can be the ratings of the users towards the watched movies in movie platforms or the stars of the videos in short-video platforms. The goal of RecSys is to maximize the long-term user satisfactory and user engagement, which is captured by the cumulative rewards.

\begin{figure*}[htbp]
\centering
\includegraphics[width=0.85\linewidth]{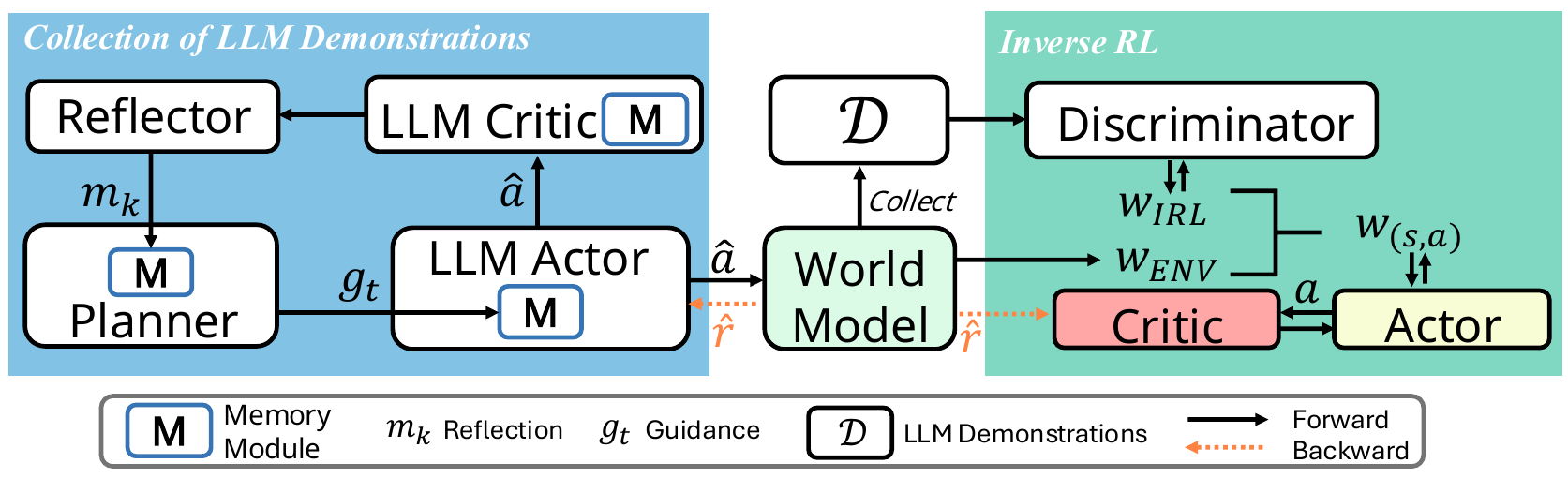}
\caption{The left part illustrates the collection of demonstrations from the LLM policy. Expert demonstrations are synthesized through the collaboration of the Planner, Reflector, LLM-Actor, and LLM-Critic. Subsequently, as shown on the right, adversarial inverse reinforcement learning (IRL) is employed to extract an effective reward function by training a discriminator to distinguish expert-generated trajectories from those produced by the current policy interacting with the learned world model. Through iterative interactions with the world model, the policy is progressively refined to surpass the original LLM expert demonstrations, while mitigating overfitting to artifacts of the simulated environment.}
\label{fig:framework}
\end{figure*}

\section{Method}
The proposed method can learn from offline expert demonstrations obtained by frozen LLM policies, which eases the need for persistent calling of LLM APIs. The collection of LLM expert demonstrations will be illustrated in the first subsection. Then, the inverse RL process is introduced to learn from the above demonstrations. 

\subsection{LLM Demonstrations}
\label{sec:demonstration}
The demonstrations are generated by LLM expert, including four modules: Reflector, Planner, LLM Actor, and LLM Critic. Prompts examples can be found in Appendix~\ref{app:prompt}.

\subsubsection{\textbf{Reflector}}
It processes completed interaction episodes to generate textual reflections, facilitating planning and decision-making. Given an episode $\mathcal{E} = \{(\mathbf{s}_t, \mathbf{a}_t, \mathbf{r}_t)\}_{t=1}^{T_k}$, the reflector generates a reflection $\mathbf{m}_k$ by a summarization function $f_\text{ref}$ implemented with LLM:
\begin{equation}
    \mathbf{m}_k = f_\text{ref}(\mathcal{E}, \mathcal{I}_r).
\end{equation}
$\mathcal{I}_r$ is the instructions for the reflector with few-shot examples. The template of reflection is provided in Fig.~\ref{fig:reflector}. The generated reflection $\mathbf{m}_k$ is stored in the memory module of the Planner, \textit{i.e.}, $\mathcal{M}_P$, and retrieved for guiding decision-making:
\begin{equation}
    \mathcal{M}_P \leftarrow \mathcal{M}_P \cup \{\mathbf{m}_k\}.
\end{equation}
The reflector is responsible for summarizing macro-level information throughout the recommendation process.

\subsubsection{\textbf{Planner}} 
It generates high-level textual plans $\mathbf{g}_t$ at each time step. 
These plans capture users' preferences and facilitate the LLM actor to execute actions. Given the user state $\mathbf{s}_t$, which is the textual description of the past interactions, \textit{i.e.}, $\mathcal{E}_t$, the planner retrieves relevant historical reflections $\mathbf{M}_s$ from the planner's memory module using Facebook AI Similarity Search (FAISS)-based~\cite{johnson2019faiss} similarity search:
\begin{equation}
    \mathcal{M}_p = \{m' \in \mathcal{M}_P | \text{sim}(m', s_t) > \tau_P\},
\end{equation}
where $\text{sim}(\cdot,\cdot)$ is the similarity function measuring the alignment between the current state and stored reflections. $\tau_P$ is the similarity threshold of reflections, after Min-Max normalization to $[0, 1]$. Subsequently, the Planner generates textual guidance $\mathbf{g}_t$ via an LLM-based planning function:
\begin{equation}
    \mathbf{g}_t = f_\text{plan}(\mathbf{s}_t, \mathcal{M}_p, \mathcal{I}_P).
\end{equation}
$\mathcal{I}_P$ denotes instructions of the Planner, including few-shot examples. Fig.~\ref{fig:planner} illustrates the structure of the prompt template. The Planner provides direct guidance for the item recommendation, enabling closer alignment with user preferences. 

\subsubsection{\textbf{LLM Actor}}
It generates executable actions at each step $t$ through a two-stage process: action generation and candidate selection. The inputs of LLM Actor include: 

(I) \textbf{Guidance:} The guidance text $\mathbf{g}_t$ from the Planner. 

(II) \textbf{Actor Memory Retrieval:} The actor retrieves relevant information from its memory module using FAISS-based similarity search:
\begin{equation}
    \mathcal{M}_a = \{m' \in \mathcal{M}_A | \text{sim}(m', \mathbf{s}_t) > \tau_A\},
\end{equation}
where $\tau_A$ is the similarity threshold after normalization. 

(III) \textbf{Instructions for LLM Actor:} $\mathcal{I}_A$ consists of a request and a few examples. The Actor function, \textit{i.e.}, $f_\text{act}$, is also implemented as a frozen LLM:
    \begin{equation}
    \label{eq:llm_actor}
        \hat{\mathbf{a}}_t = f_\text{act}(\mathbf{s}_t, \mathbf{g}_t, \mathcal{M}_a, \mathcal{I}_A).
    \end{equation}
For ease of understanding, an illustrative prompt template of LLM-actor is provided in Fig.~\ref{fig:actor}. 

Since the generated $\hat{\mathbf{a}}_t$ may not exactly coincide with any item embedding, it is regarded as an indicator of the candidate set, and the item with the highest cosine similarity is chosen as the final action:
\begin{equation}
     a_t = \arg\max_{\mathcal{I}} \sum_{i \in \mathcal{I}} \text{sim}(\hat{\mathbf{a}}_t, \mathbf{e}_i),
\end{equation}
where $\mathcal{I}$ is item set and $\mathbf{e}_i$ stands for item embedding.

After each recommendation, the Actor updates its memory with the new state-action pair for future retrieval.

\subsubsection{\textbf{LLM Critic}} 
It estimates the value function to evaluate the actor’s action, and the prompt template is shown in Fig.~\ref{fig:critic}. It first retrieves relevant historical state-value pairs from its memory module:
\begin{equation}
    \mathcal{M}_c = \{m' \in \mathcal{M}_C | \text{sim}(m', \mathbf{s}_t) > \tau_C\},
\end{equation}
where $\tau_C$ is the similarity threshold. The retrieved information, combined with the critic’s instructions, \textit{i.e.}, $\mathcal{I}_C$, is fed into the LLM-based critic to estimate the state value:
\begin{equation}
\label{eq:llm_critic}
    V_{\text{LLM}}(s_t) = f_\text{cri}(\mathbf{s}_t, \mathcal{M}_c, \mathcal{I}_C).
\end{equation}

To stabilize training, the advantage function is introduced following Advantage Actor Critic (A2C)~\cite{mnih2016a2c}:
\begin{equation}
\label{eq:llm_adv}
    A(s_t, a_t) = Q(s_t, a_t) - V(s_t),
\end{equation}
where $Q(s_t, a_t) = r_t + \gamma V_{\text{LLM}}(s_{t+1})$ is also obtained from the estimated value function of the LLM Critic. At the end of each step, the Critic updates the memory modules with the new state-value pairs.

Through the interaction between the world model and the LLM modules, the LLM expert generates expert demonstrations $\mathcal{D}_E$, structured as sequential state-action trajectories:
\begin{equation}
\label{eq:demo}
    \mathcal{D}_E = \{\tau_u\}_{u=1}^{U},
    \tau_u = \{(s_1,a_1),(s_2,a_2),\dots,(s_{T_u},a_{T_u})\}.
\end{equation}
Each state-action pair is encoded into textual embeddings using the LLaMA~\cite{touvron2023llama2} model, ensuring semantically coherent and expressive representations for subsequent AIRL, bridging the expert policy generation and imitation learning.

\subsection{Adversarial Inverse RL}
In order to extract a reward function from the LLM-generated demonstrations, an adversarial inverse reinforcement learning (IRL) procedure is employed. A discriminator \(D_\psi\) is introduced to distinguish between the trajectories generated by the current learning policy \(\pi_\theta\) and the expert demonstrations \(\mathcal{D}_E\). 
It is interpreted as the probability that \((s,a)\) originates from the policy rather than from the expert.  

During the update of the discriminator, trajectories \(\tau_\theta\) are sampled by rolling out \(\pi_\theta\) in the learned world model \(\hat{T}\). Simultaneously, batches of state–action pairs are drawn from \(\mathcal{D}_E\). The discriminator parameters \(\psi\) are then optimized to minimize the cross-entropy loss, similar to Eq.~\ref{eq:obj}:
\begin{equation}
\begin{split}
    \mathcal{L}_D(\psi) & =\mathbb{E}_{(s,a)\sim \rho_{\mathcal{D}_E}} \bigl[\log\bigl(1 - D_{\psi}(s,a)\bigr)\bigr] \\ &
    -\mathbb{E}_{(s,a)\sim \rho_{\pi_\theta}} \bigl[\log D_{\psi}(s,a)\bigr].
\end{split}
\end{equation}
Correspondingly, the reward function is defined as:
\begin{equation}
    r_{\text{IRL}}(s,a) \;=\; -\,\log D_{\psi}(s,a).
\end{equation}
Therefore, transitions that are more likely to be classified as “expert” yield a higher reward.  

\subsubsection{\textbf{Demonstration Weighting}}
Since the LLM-generated demonstrations \(\mathcal{D}_E\) are known to be suboptimal, a weighting scheme is introduced to emphasize high-quality transitions while de-emphasizing those that are likely to be suboptimal. First, for each demonstration transition \((s,a)\), the advantage \(A_{\text{demo}}(s,a)\) is estimated using the world model's reward function, \textit{i.e.,} \(\hat{r}\),
and a value network \(V_{\text{demo}}\). Specifically, given a demonstration trajectory \(\tau = \{(s_t, a_t, r_t)\}_{t=1}^T\), the return,
\begin{equation}
    G_t \;=\; \sum_{k=t}^T \gamma^{\,k-t} \, r_k,
\end{equation}
is computed via rollouts in the world model. \(V_{\text{demo}}(s_t)\) is then estimated by fitting a neural network to minimize the mean-squared error between the predicted value and the empirically observed return. The advantage is defined as:
\begin{equation}
    A_{\text{demo}}(s_t,a_t) \;=\; Q_{\text{demo}}(s_t,a_t) \;-\; V_{\text{demo}}(s_t).
\end{equation}

Next, an environment-based weight can be computed as:
\begin{equation}
    w_{\text{env}}(s,a) \;=\; \exp\!\bigl(\tfrac{1}{\beta}\,A_{\text{demo}}(s,a)\bigr),
\end{equation}
where \(\beta\) is a temperature hyperparameter. Transitions with positive advantage are thereby assigned larger weight.  

In parallel, an IRL-based weight is defined by:
\begin{equation}
    w_{\text{IRL}}(s,a) \;=\; \Bigl(\tfrac{1}{D_{\psi}(s,a)} - 1 \Bigr)^{\gamma},
\end{equation}
where \(\gamma\) controls the confidence of the discriminator. Transitions with smaller \(D_{\psi}(s,a)\) (i.e., the discriminator is confident that \((s,a)\) resembles the expert) receive a higher \(w_{\text{IRL}}\).  

Finally, the two weights are combined into a single weight:
\begin{equation}
    w(s,a) \;=\; \bigl[w_{\text{env}}(s,a)\bigr]^{\alpha} \;\times\; \bigl[w_{\text{IRL}}(s,a)\bigr]^{\,1-\alpha},
\end{equation}
with \(\alpha \in [0,1]\) controlling the trade-off. The combined weight \(w(s,a)\) is normalized across the demonstration dataset and clipped to a predefined range to ensure numerical stability.  

\subsubsection{\textbf{Policy Update}}

At each iteration, the policy \(\pi_\theta\) is updated by combining two objectives:

\begin{enumerate}[label=(\Roman*)]
    \item \textbf{Weighted Imitation Loss:} For demonstration transitions \((s,a)\in\mathcal{D}_E\), the imitation objective is defined as:
    \begin{equation}
        \mathcal{L}_{\text{imit}}(\theta) \;=\; -\,\mathbb{E}_{(s,a)\sim \mathcal{D}_E} \bigl[\,w(s,a)\,\log \pi_{\theta}(a\mid s)\bigr].
    \end{equation}
    This loss encourages the policy to imitate demonstration transitions in proportion to their weights.
    
    \item \textbf{Reinforcement Loss (Actor–Critic):} Let \(Q_\phi\) and \(V_\phi\) denote the critic networks for estimating the environment return under \(\pi_\theta\). The actor–critic loss is given by:
    \begin{equation}
    \begin{split}
        \mathcal{L}_{\text{RL}}(\theta) = & -\mathbb{E}_{s \sim \mathcal{D}} \Bigl[\mathbb{E}_{a\sim \pi_{\theta}} \bigl[Q_\phi(s,a)\bigr] \\& - \alpha_{\text{ent}} \mathcal{H}\bigl(\pi_{\theta}(\cdot\mid s)\bigr)\Bigr],
    \end{split}
    \end{equation}
    where \(\mathcal{H}(\pi_{\theta}(\cdot\mid s))\) is the policy entropy and \(\alpha_{\text{ent}}\) is an entropy temperature. Samples are drawn from a replay buffer \(\mathcal{D}\) containing both environment-generated transitions and demonstration transitions (sampled with priority proportional to \(w(s,a)\)).  
\end{enumerate}

The total policy loss is defined as:
\begin{equation}
\label{eq:loss_policy}
    \mathcal{L}_{\text{policy}}(\theta) \;=\; \lambda_{\text{imit}}\,\mathcal{L}_{\text{imit}}(\theta) \;+\; \mathcal{L}_{\text{RL}}(\theta),
\end{equation}
where \(\lambda_{\mathrm{imit}}\) balances imitation and reinforcement learning. The critic parameters \(\phi\) are updated using temporal-difference targets, while the discriminator \(D_\psi\) is periodically retrained with updated rollouts and demonstrations.

\subsection{Overall Learning Pipeline}
The complete training procedure, as detailed in Fig.~\ref{fig:framework} consists of three phases: (i) world model learning, (ii) demonstration processing and weighting, and (iii) policy optimization.  

\subsubsection{\textbf{World Model Learning}}
Similar to existing model-based methods~\cite{gao2023dorl,yi2024roler}, offline logs are utilized to learn the reward function, \textit{i.e.}, $\hat{r}$, of the world model via supervised learning, such as DeepFM~\cite{deepfm}. The transition function is handled by a state tracker, which is implemented as a sequential model, encoding historical interactions as the next state.

\subsubsection{\textbf{Demonstration Processing and Weighting}}
Then, the LLM expert demonstrations \(\mathcal{D}_E\) are collected as described in Section~\ref{sec:demonstration}. For each transition \((s,a)\in \mathcal{D}_E\):
\begin{enumerate}[label=(\Roman*)]
    \item Advantage \(A_{\text{demo}}(s,a)\) is computed via rollouts in the world model and a dedicated value network \(V_{\text{demo}}\).  
    \item The discriminator \(D_{\psi}\) is trained on \(\mathcal{D}_E\) and on policy rollouts \(\rho_{\pi_\theta}\) in the world model.  
    \item Weights \(w_{\text{env}}(s,a)\) and \(w_{\text{IRL}}(s,a)\) are computed, and fused to obtain \(w(s,a)\). These weights are then normalized and stored in a replay buffer \(\mathcal{B}_{\text{demo}}\).
\end{enumerate}
  
\subsubsection{\textbf{Policy Optimization}}
At last, the policy \(\pi_\theta\) and critics \((Q_\phi, V_\phi)\) are updated using both environment rollouts and weighted demonstration samples:
\begin{enumerate}[label=(\Roman*)]
    \item A batch of transitions \(\{(s_i,a_i,r_i,s'_i)\}\) is sampled from a mixed replay buffer \(\mathcal{B} = \mathcal{B}_{\text{env}} \cup \mathcal{B}_{\text{demo}}\), where \(\mathcal{B}_{\text{env}}\) contains recent rollouts of \(\pi_\theta\), and \(\mathcal{B}_{\text{demo}}\) contains demonstration transitions paired with weights \(w(s,a)\).  
    \item Critic networks are updated by minimizing the temporal-difference loss:
    \begin{equation}
        \mathcal{L}_{\text{critic}}(\phi) \;=\; \mathbb{E}_{(s,a,r,s')\sim \mathcal{B}} \Bigl[\bigl(Q_{\phi}(s,a) - y\bigr)^2\Bigr],
    \end{equation}
    where the target $ y = r + \gamma \,V_{\phi_{\text{target}}}(s')$, and \(\phi_{\text{target}}\) denotes delayed parameters. Demonstration samples contribute to \(\mathcal{L}_{\text{critic}}\) in proportion to their weights \(w(s,a)\).  
    \item The policy is updated by minimizing \(\mathcal{L}_{\text{policy}}(\theta)\) in Eq.~\ref{eq:loss_policy}. 
    \item Periodically, the discriminator \(D_{\psi}\) is retrained using the updated demonstration weights and rollout data from \(\pi_\theta\).  
\end{enumerate}

The last two phases are repeated until convergence. By iteratively re-estimating demonstration weights and updating the policy via actor–critic and weighted imitation, the learned policy is driven to exceed the performance of the original LLM demonstrations, avoiding overfitting to simulator artifacts.

\section{Experiments}
In this section, extensive experiments are conducted to validate the effectiveness of the proposed method by answering the following research questions (RQs):

\noindent$\bullet$ (RQ1) How is the recommendation performance of IL-Rec compared to other baselines?

\noindent$\bullet$ (RQ2) How do the components and designs of IL-Rec contribute to its performance?

\noindent$\bullet$ (RQ3) How does the generalization capacity of LLMs influence the performance of IL-Rec?

\noindent$\bullet$ (RQ4) How is the robustness of IL-Rec against its critical hyperparameters? 

\begin{table}[!t]
    \centering
    \caption{Dataset statistics.}
    \vspace{-1em}
    \label{tb:dataset}
    \setlength{\tabcolsep}{10pt}
    \setlength{\aboverulesep}{1pt}
    \setlength{\belowrulesep}{1pt}
    \resizebox{0.95\columnwidth}{!}{
        \begin{tabular}{ccccc}
            \toprule
            Dataset                   & Usage & \# Users & \# Items & \makecell{Density}   \\
            \toprule
            \multirow{2}{*}{Steam} & Train & 6012        &  190365         & 0.145\%              \\
                                  & Test  & 6012        &  190365         & 0.084\%             \\
            \midrule
            \multirow{2}{*}{Amazon}  & Train &  3109         &  13864        & 0.788\%             \\
                                  & Test  & 3109         &  13864         & 0.320\%             \\
            \bottomrule
        \end{tabular}
    }
    \vspace{-0.8em}
\end{table}

\subsection{Setup}
\subsubsection{\textbf{Datasets}}
Experiments are conducted on two benchmark datasets following BiLLP~\cite{shi2024billp} with statistics in Tab.~\ref{tb:dataset}.

\noindent $\bullet$ \textbf{Steam}~\cite{kang2018sasrec} contains user histories\footnote{\url{https://github.com/jizhi-zhang/BiLLP}} from a game platform. The rating of a user toward a game is measured by the game time, ranging from 1 to 5, serving as the immediate rewards.

\noindent $\bullet$ \textbf{Amazon}~\cite{ni2019amazon-book} contains book rating histories\footnotemark[1]. The genres and titles of books serve as the text information. The range of ratings is also $[1, 5]$ and serves as an immediate rewards.

\begin{table*}
  \centering
  \caption{The performance summary of all methods on Steam and Amazon. (Bold: best; Underline: runner-up)}
  \vspace{-0.5em}
  \label{tab:steam_amazon}
  \setlength{\aboverulesep}{1pt}
  \setlength{\belowrulesep}{1pt}
  \resizebox{0.9\textwidth}{!}{
      \begin{tabular}{l|ccc|ccc}
        \toprule[1pt]
        & \multicolumn{3}{c|}{Steam} & \multicolumn{3}{c}{Amazon} \\
        \cmidrule(lr){2-4}\cmidrule(lr){5-7}
        Method & Len & $R_{\text{each}}$ & $R_{\text{traj}}$ & Len & $R_{\text{each}}$ & $R_{\text{traj}}$ \\
        \midrule[1pt]
        SQN      & 2.183 $\pm$ 0.177 & 3.130 $\pm$ 0.050 &  6.837 $\pm$ 0.517 & 4.773 $\pm$ 0.059 & 4.303 $\pm$ 0.017 & 20.570 $\pm$ 0.245 \\
        CRR      & 4.407 $\pm$ 0.088 & 3.263 $\pm$ 0.427 & 14.377 $\pm$ 1.568 & 3.923 $\pm$ 0.162 & 4.537 $\pm$ 0.130 & 17.833 $\pm$ 1.129 \\
        BCQ      & 4.720 $\pm$ 0.343 & 3.997 $\pm$ 0.068 & 18.873 $\pm$ 1.092 & 4.847 $\pm$ 0.721 & 4.367 $\pm$ 0.053 & 21.150 $\pm$ 2.893 \\
        CQL      & 5.853 $\pm$ 0.232 & 3.743 $\pm$ 0.147 & 21.907 $\pm$ 0.299 & 2.280 $\pm$ 0.185 & 4.497 $\pm$ 0.039 & 10.263 $\pm$ 0.882 \\
        DQN      & 4.543 $\pm$ 0.693 & 4.500 $\pm$ 0.069 & 20.523 $\pm$ 3.618 & 4.647 $\pm$ 0.498 & 4.290 $\pm$ 0.083 & 19.923 $\pm$ 1.909 \\
        A2C      & 9.647 $\pm$ 0.848 & 4.367 $\pm$ 0.069 & 42.180 $\pm$ 3.937 & 7.873 $\pm$ 0.310 & 4.497 $\pm$ 0.026 & 35.437 $\pm$ 1.453 \\
        DORL     & 9.467 $\pm$ 0.862 & 4.033 $\pm$ 0.098 & 38.300 $\pm$ 4.173 & 7.507 $\pm$ 0.174 & 4.510 $\pm$ 0.014 & 33.887 $\pm$ 0.655 \\
        ROLeR    & 10.218 $\pm$ 0.783 & 4.208 $\pm$ 0.086 & 42.997 $\pm$ 3.448 & 8.604 $\pm$ 0.243 & 4.590 $\pm$ 0.046 & 39.492 $\pm$ 0.632 \\
        \midrule[0.5pt]
        ActOnly  & 5.567 $\pm$ 0.160 & \underline{$4.537 \pm 0.021$} & 25.250 $\pm$ 0.637 & 6.383 $\pm$ 0.176 & 4.490 $\pm$ 0.008 & 28.660 $\pm$ 0.761 \\
        ReAct    & 11.630 $\pm$ 0.741 & $\mathbf{4.559 \pm 0.047}$ & 42.990 $\pm$ 2.925 & 7.733 $\pm$ 0.450 & \underline{4.603 $\pm$ 0.033} & 35.063 $\pm$ 1.106 \\
        Reflexion & 12.690 $\pm$ 1.976 & 4.523 $\pm$ 0.026 & 57.423 $\pm$ 8.734 & 8.700 $\pm$ 0.535 & $\mathbf{4.670 \pm 0.073}$ & 40.470 $\pm$ 2.954 \\
        BiLLP    & \underline{15.367 $\pm$ 0.119} & 4.503 $\pm$ 0.069 & \underline{69.193 $\pm$ 1.590} & \underline{9.413 $\pm$ 0.190} & 4.507 $\pm$ 0.012 & \underline{42.443 $\pm$ 0.817} \\
        \midrule[0.5pt]
        \rowcolor{gray!20} IL-Rec (Ours)     & $\mathbf{17.533 \pm 2.602}$ & 4.476 $\pm$ 0.324 & $\mathbf{78.478 \pm 3.267}$ &
        $\mathbf{11.136 \pm 0.797}$ & 4.330 $\pm$ 0.114 & $\mathbf{48.219 \pm 2.198}$ \\
        \bottomrule[1pt]
      \end{tabular}
  }
\vspace{-0.5em}
\end{table*}

\subsubsection{\textbf{Baselines}}
The baseline methods can be classified into two types: offline reinforcement learning methods and LLM methods. For all model-based methods, similar to prior works~\cite{gao2023cirs,gao2023dorl,yi2024roler}, world models are trained through DeepFM~\cite{deepfm}, which means they share the same world models for fair comparison of their recommendation performance.

\noindent \textbf{Offline RL}:

$\bullet$ \textbf{\textit{DQN} (2015)}~\cite{mnih2015dqn} is a fundamental Q-learning-based method that employs deep neural networks to estimate the value function.

$\bullet$ \textbf{\textit{A2C} (2016)}~\cite{mnih2016a2c} is a fundamental actor-critic baseline which exploits the advantage function to stabilize training.

$\bullet$ \textbf{\textit{SQN (2020)}}~\cite{xin2020sqn} employs a dual-head architecture, comprising a cross-entropy loss head and an RL-based head. The RL head is in charge of generating the final recommendations.

$\bullet$ \textbf{\textit{CRR (2020)}}~\cite{wang2020crr} is a model-free RL method that focuses on improving policy learning by explicitly avoiding out-of-distribution (OOD) actions, enhancing stability and robustness.

$\bullet$ \textbf{\textit{BCQ (2019)}}~\cite{fujimoto2019bcq} is an extension of deep Q-learning tailored for batch RL scenarios. It integrates a discrete-action mechanism that filters out uncertain data points, ensuring that policy updates rely only on high-confidence samples.

$\bullet$ \textbf{\textit{CQL (2020)}}~\cite{kumar2020cql} is another model-free RL approach. It introduces a Q-value regularization mechanism within an actor-critic framework, aiming to constrain the learned policy and mitigate overestimation issues.

$\bullet$ \textbf{\textit{DORL (2023)}}~\cite{gao2023dorl} is a model-based RL method which serves as a strong baseline on KuaiRec~\cite{gao2022kuairec} and KuaiRand~\cite{gao2022kuairand}. It reconstructs the uncertainty penalty and introduces an entropy penalty to ease the Mathew effect.

$\bullet$ \textbf{\textit{ROLeR (2024)}}~\cite{yi2024roler} is a non-parametric soft-label-based reward shaping method. It designs an uncertainty penalty tailored for model-based offline RL4RS.

\noindent \textbf{LLM Methods}:

$\bullet$ \textbf{\textit{ActOnly}} serves as a naive LLM baseline which directly recommends items with LLM Actor-Critic.

$\bullet$ \textbf{\textit{ReAct (2023)}}~\cite{yao2023react} integrates LLMs to simultaneously generate reasoning traces and execute task-specific actions in an interleaved fashion. By combining reasoning and decision-making within the same framework, it enhances the coherence between thought processes and actions.

$\bullet$ \textbf{\textit{Reflextion (2023)}}~\cite{shinn2024reflexion} employs verbalized self-reflection based on task feedback, storing these reflections in an episodic memory buffer. The retained reflective insights improve decision-making in future interactions by allowing the model to refine its responses iteratively.

$\bullet$ \textbf{\textit{BiLLP (2024)}}~\cite{shi2024billp} is the current SOTA on Steam and Amazon-Book with respect to cumulative reward evaluation. It exploits the LLM Actor-Critic as a direct recommendation policy. The update of its policy relies on the update of the memory modules.

\subsubsection{\textbf{Evaluation and Metric}}
To systematically evaluate the effectiveness of IL-Rec, a structured evaluation protocol consistent with prior studies~\cite{gao2023cirs,gao2023dorl,yi2024roler,shi2024billp} is retained. The evaluation focuses on key performance indicators that assess both short-term and long-term user satisfaction.

\noindent $\bullet$ $R_\text{tra}$ measures the average cumulative reward over testing episodes. As it directly reflects long-term user engagement and satisfaction, it serves as the primary metric for evaluation.

\noindent $\bullet$ $R_\text{each}$ denotes the mean single-step reward obtained during testing. This metric captures the immediate satisfaction level of users with each recommended item.

\noindent $\bullet$ Len represents the average interaction length across test trajectories, offering insights into the recommender system’s ability to sustain user engagement.

The recommendation session is terminated under two conditions: (1) If, within the most recent $N$ transitions, a user has received $M$ items from the same category, the session ends. To ensure fair comparison with~\cite{shi2024billp}, $N=50, M=4$ for the Steam dataset and $M=4, N=15$ for Amazon is adopted, enforcing category diversity within every four interactions. (2) The session also ends when the interaction length reaches the predefined upper limit of 100.

\subsubsection{\textbf{Implementation}}
\label{sec:implementation}
The choices of hyperparameters are summarized at here for readability.
The temperature of the LLMs is set to $0.5$ to ensure the flexibility of the generation. The temperature $\beta$ in $w_{\mathrm{env}}$ is chosen from $\{0.1,\,1.0,\,10.0\}$. The trade‐off coefficient $\alpha$ in weight fusion is turned in $\{0.25,\,0.5,\,0.75\}$. The entropy temperature $\alpha_{\mathrm{ent}}$ is chosen from $\{0.01,\,0.1,\,0.2\}$. The imitation loss weight $\lambda_{\mathrm{imit}}$ is tuned over $\{0.1,\,0.5,\,2.0\}$.
Adam is the default optimizer used for all the components of the RL Actor-Critic in inverse RL. The learning rate is set to $0.001$.
A single trial of IL-Rec requires about 12 GPU-hours on an NVIDIA Tesla A100 (40 GB HBM2), consisting of 3 GPU-hours for world model simulation, 3 GPU-hours for generating LLM demonstrations, and 6 GPU-hours for adversarial inverse RL. In comparison, BiLLP consumes 6 GPU-hours per trial. While IL-Rec roughly doubles the computational cost, it delivers substantial gains in long-term user satisfaction. Moreover, both the world model simulation and the demonstration generation are one-time costs, further illustrating the practicality of IL-Rec.

\subsection{Overall Performance (RQ1)}
To evaluate the effectiveness of IL-Rec, its performance is compared against a suite of offline RL and LLM-based baselines. Tab.~\ref{tab:steam_amazon} reports the results on the Steam and Amazon datasets.

\textbf{Cumulative Reward (\(R_{\mathrm{traj}}\)).}  
IL-Rec is observed to attain the highest cumulative reward on both benchmarks. On Steam, IL-Rec achieves $78.478$, surpassing the previous best, \textit{i.e.,}, BiLLP, by 13.4\%. On Amazon, it achieves $48.219$, an improvement of 13.6\% over BiLLP. These gains demonstrate that the proposed imitation-enhanced RL framework is able to learn policies that substantially improve long-term engagement.

\textbf{Single-Step Reward (\(R_{\mathrm{each}}\)).}  
Competitive single-step rewards are produced by IL-Rec, $4.476$ on Steam and $4.330$ on Amazon, although it does not hold the top position. Methods based on direct LLM reasoning—ReAct and ActOnly —yield slightly higher immediate rewards, as their designs prioritize short-term engagement. However, their cumulative rewards remain considerably lower, underscoring the advantage of balancing immediate gains with long-term strategy via RL.

\textbf{Interaction Length} (Len)\textbf{.}
The longest recommendation trajectories are produced by IL-Rec, indicating enhanced sustained engagement. On Steam, the average trajectory length of $17.533$ represents a 14.1\% increase over BiLLP’s. On Amazon, the length of $11.136$ corresponds to an 18.3\% increase over the second best. This extension of interaction horizons is attributed to the integration of structured demonstration weighting and RL-driven policy refinement.

\textbf{Baseline Analysis.}  
Among the RL-based baselines, A2C, DORL, and ROLeR produce higher rewards and longer trajectories than value-based methods (DQN, BCQ, CQL), yet they remain inferior to IL-Rec, as they lack demonstration weighting and knowledge-guided imitation. Conversely, LLM-based methods such as BiLLP, ReAct, and Reflexion exhibit strong one-step reasoning but fail to optimize long-term returns due to their static policy updates. Overall, IL-Rec is shown to successfully bridge the gap between RL-based and LLM-based recommendation strategies, achieving superior long-term performance while retaining competitive immediate rewards.  

\subsection{Ablation Study (RQ2)}
This section examines the contributions of each component within IL-Rec. Variants are defined as follows: “w/o \(w\)” sets \(w(s,a)=1\); “w/o \(w_{\mathrm{env}}\)” uses only the IRL-based weight; “w/o \(w_{\mathrm{IRL}}\)” uses only the environment-based weight; and “BiLLP” corresponds to disabling the entire adversarial IRL procedure. The results on Steam and Amazon are summarized in Tab.~\ref{tab:ablation}.

\textbf{Impact of Overall Weighting.}  
When the weighting mechanism was removed (w/o \(w\)), cumulative reward (\(R_{\mathrm{traj}}\)) is reduced 14.1\% on Steam and 18.4\% on Amazon. Interaction length (Len) likewise decreased by 14.5\% and 22.5\%, respectively. Single-step reward (\(R_{\mathrm{each}}\)) remained effectively unchanged. These findings indicate that the weighting mechanism is essential for achieving long-horizon engagement.

\textbf{Impact of Environment-Based Weight (\(w_{\mathrm{env}}\)).}  
When \(w_{\mathrm{env}}\) is ablated, cumulative reward fell to \(63.380\) on Steam (–19.3\%) and to \(40.280\) on Amazon (–16.5\%), while Len dropped by 24.7\% and 22.3\%, respectively. In contrast, \(R_{\mathrm{each}}\) increased by 7.2\% and 7.6\%, respectively. This indicates that environment-based advantage weighting is critical for optimizing long-term returns, at the cost of slight increases in immediate reward.

\textbf{Impact of IRL-Based Weight (\(w_{\mathrm{IRL}}\)).}  
When \(w_{\mathrm{IRL}}\) was removed, cumulative reward decreased to \(71.350\) on Steam (–9.1\%) and to \(44.000\) on Amazon (–8.7\%), with Len reductions of 10.6\% and 1.0\%. Single-step reward changed marginally (+1.7\% on Steam, –7.9\% on Amazon). These results confirm that IRL-based weighting principally supports long-horizon performance, exerting only minor influence on immediate rewards.

\textbf{Comparison with BiLLP.}  
On Steam and Amazon, the recommendation policies learned by the BiLLP baseline are not as effective as IL-Rec in sustaining long-term user satisfaction. Interaction lengths and \(R_{\mathrm{each}}\) values under BiLLP further highlight that IL-Rec’s dual-weighting and RL integration are responsible for its superior long-term engagement and competitive single-step performance. In addition, the results also show that without the capacity of learning from sub-optimal demonstrations, purely imitating the LLM polices cannot learn effective recommendation policies.

\begin{table}[t]
    \centering
    \caption{Ablation Study on IL-Rec's components and designs.}
    \vspace{-0.5em}
    \begin{minipage}[t]{\columnwidth}
        \setlength{\tabcolsep}{7.5pt}
        \setlength{\aboverulesep}{1pt}
        \setlength{\belowrulesep}{1pt}
        \resizebox{\textwidth}{!}{
            \begin{tabular}{lccc}
                \toprule
                \multirow{2}{*}{Methods} & \multicolumn{3}{c}{Steam} \\
                \cmidrule(lr){2-4}
                & Len & $R_{\text{each}}$ & $R_{\text{traj}}$ \\
                \midrule
                BiLLP     & 15.367 $\pm$ 0.119 & 4.503 $\pm$ 0.069 & 69.193 $\pm$ 1.590 \\
                w/o $w$   & 14.990 $\pm$ 1.264 & 4.495 $\pm$ 0.664 & 67.380 $\pm$ 1.780 \\
                w/o $w_{\mathrm{env}}$   & 13.210 $\pm$ 1.460 & 4.798 $\pm$ 0.394 & 63.380 $\pm$ 2.142 \\
                w/o $w_{\mathrm{IRL}}$    & 15.670 $\pm$ 1.297 & 4.553 $\pm$ 0.535 & 71.350 $\pm$ 1.272 \\
                IL-Rec      & 17.533 $\pm$ 2.602 & 4.476 $\pm$ 0.324 & 78.478 $\pm$ 3.267 \\
                \bottomrule
            \end{tabular}
        }
    \end{minipage}
    \begin{minipage}[t]{\columnwidth}
        \setlength{\tabcolsep}{7.5pt}
        \setlength{\aboverulesep}{1pt}
        \setlength{\belowrulesep}{1pt}
        \resizebox{\textwidth}{!}{
            \begin{tabular}{lccc}
                \toprule
                \multirow{2}{*}{Methods} & \multicolumn{3}{c}{Amazon} \\
                \cmidrule(lr){2-4}
                & Len & $R_{\text{each}}$ & $R_{\text{traj}}$ \\
                \midrule
                BiLLP     & 9.413 $\pm$ 0.190 & 4.507 $\pm$ 0.012 & 42.443 $\pm$ 0.817 \\
                w/o $w$   & 8.630 $\pm$ 0.663 & 4.560 $\pm$ 0.056 & 39.350 $\pm$ 1.225 \\
                w/o $w_{\mathrm{env}}$   & 8.650 $\pm$ 0.723 & 4.657 $\pm$ 0.399 & 40.280 $\pm$ 1.450 \\
                w/o $w_{\mathrm{IRL}}$    & 11.030 $\pm$ 0.713 & 3.989 $\pm$ 0.058 & 44.000 $\pm$ 1.256 \\
                IL-Rec      & 11.136 $\pm$ 0.797 & 4.330 $\pm$ 0.114 & 48.219 $\pm$ 2.198 \\
                \bottomrule
            \end{tabular}
        }
    \end{minipage}
    \label{tab:ablation}
\end{table}

\begin{table}[t]
    \centering
    \caption{IL-Rec's Performance w/ Different LLMs.}
    \vspace{-0.5em}
    \begin{minipage}[t]{\columnwidth}
        \setlength{\tabcolsep}{3pt}
        \setlength{\aboverulesep}{1pt}
        \setlength{\belowrulesep}{1pt}
        \resizebox{\textwidth}{!}{
            \begin{tabular}{lccc}
                \toprule
                \multirow{2}{*}{Methods} & \multicolumn{3}{c}{Steam} \\
                \cmidrule(lr){2-4}
                & Len & $R_{\text{each}}$ & $R_{\text{traj}}$ \\
                \midrule
                Llama2-7b         & 16.900 $\pm$ 1.296 & 4.246 $\pm$ 0.213 & 71.750 $\pm$ 1.611 \\
                Deepseek-R1-14b   & 17.600 $\pm$ 1.119 & 4.250 $\pm$ 0.134 & 74.800 $\pm$ 2.496 \\
                GPT3.5-16k   & 17.533 $\pm$ 2.602 & 4.476 $\pm$ 0.324 & 78.478 $\pm$ 3.267 \\
                GPT4-32k    & 27.100 $\pm$ 2.345 & 4.736 $\pm$ 0.286 & 128.346 $\pm$ 1.671 \\
                \bottomrule
            \end{tabular}
        }
    \end{minipage}
    \begin{minipage}[t]{\columnwidth}
        \setlength{\tabcolsep}{2.5pt}
        \setlength{\aboverulesep}{1pt}
        \setlength{\belowrulesep}{1pt}
        \resizebox{\textwidth}{!}{
            \begin{tabular}{lccc}
                \toprule
                \multirow{2}{*}{Methods} & \multicolumn{3}{c}{Amazon} \\
                \cmidrule(lr){2-4}
                & Len & $R_{\text{each}}$ & $R_{\text{traj}}$ \\
                \midrule
                Llama2-7b     & 9.820 $\pm$ 0.583 & 4.122 $\pm$ 0.593 & 40.480 $\pm$ 1.783 \\
                Deepseek-R1-14b  & 10.770 $\pm$ 0.688 & 4.160 $\pm$ 0.088 & 44.800 $\pm$ 1.241 \\
                GPT3.5-16k   & 11.136 $\pm$ 0.797 & 4.330 $\pm$ 0.114 & 48.219 $\pm$ 2.198 \\
                GPT4-32k    & 13.270 $\pm$ 0.552 & 4.832 $\pm$ 0.069 & 64.121 $\pm$ 1.064 \\
                \bottomrule
            \end{tabular}
        }
    \end{minipage}
    \label{tab:scaling}
\end{table}

\subsection{Impact of LLM Generalization (RQ3)}
To evaluate how the performance of IL-Rec varies with different large language models (LLMs), experiments are conducted with Llama2-7b-hf-chat, Deepseek-R1-Distill-Qwen-14b, GPT-3.5-turbo-16k, and GPT-4-32k. Tab.~\ref{tab:scaling} presents the results on the Steam and Amazon datasets.

\textbf{Scaling LLMs Improves Performance.} The results indicate that as the LLM’s capability increases, IL-Rec consistently benefits from the enhanced reasoning and generalization abilities. Using GPT-4-32k leads to the highest cumulative reward ($R_\text{tra}$), achieving $128.346$ on Steam and $64.121$ on Amazon, which significantly outperforms the other variants. Additionally, the interaction length (Len) also sees a substantial increase, confirming that stronger LLMs contribute to prolonged engagement and better policy learning.

\textbf{Performance with Smaller LLMs.} Even when using smaller-scale LLMs such as Llama2-7b and Deepseek-R1-14b, IL-Rec achieves competitive results compared to BiLLP. Notably, IL-Rec with Llama-7b attains $71.75$ in $R_\text{tra}$ on Steam, which is comparable to BiLLP’s performance. Similarly, the results of IL-Rec with Deepseek-R1-14b have already surpassed BiLLP’s cumulative reward on Amazon. This suggests that IL-Rec can effectively leverage even moderately capable LLMs to achieve strong recommendation policies, reducing reliance on high-resource models.

\textbf{Implications for Future Improvements.} The significant performance gains observed with more powerful LLMs suggest that IL-Rec has room for further improvement as LLM architectures continue to advance. The increasing gap in $R_\text{tra}$ between Llama2-7b and GPT-4-32k implies that better LLM generalization leads to better long-term decision-making. This trend reinforces the potential of future iterations of IL-Rec incorporating more sophisticated LLMs with improved retrieval augmentation and decision planning.

Policies learned from LLM demonstrations surpass the static LLM experts, confirming the value of combining demonstrations with reinforcement learning. Yet, the suboptimality of these demonstrations is only partially alleviated. The extent of improvement depends on factors such as the reliability of the weighting mechanism, demonstration coverage, and biases. These factors delineate the capability boundary of IL-Rec, and further extension will be investigated in future work.
\vspace{-0.3em}

\subsection{Hyperparameter Sensitivity (RQ4)}
\begin{figure}[!t]
    \centering
    \begin{minipage}[t]{0.485\columnwidth}
        \includegraphics[width=\columnwidth]{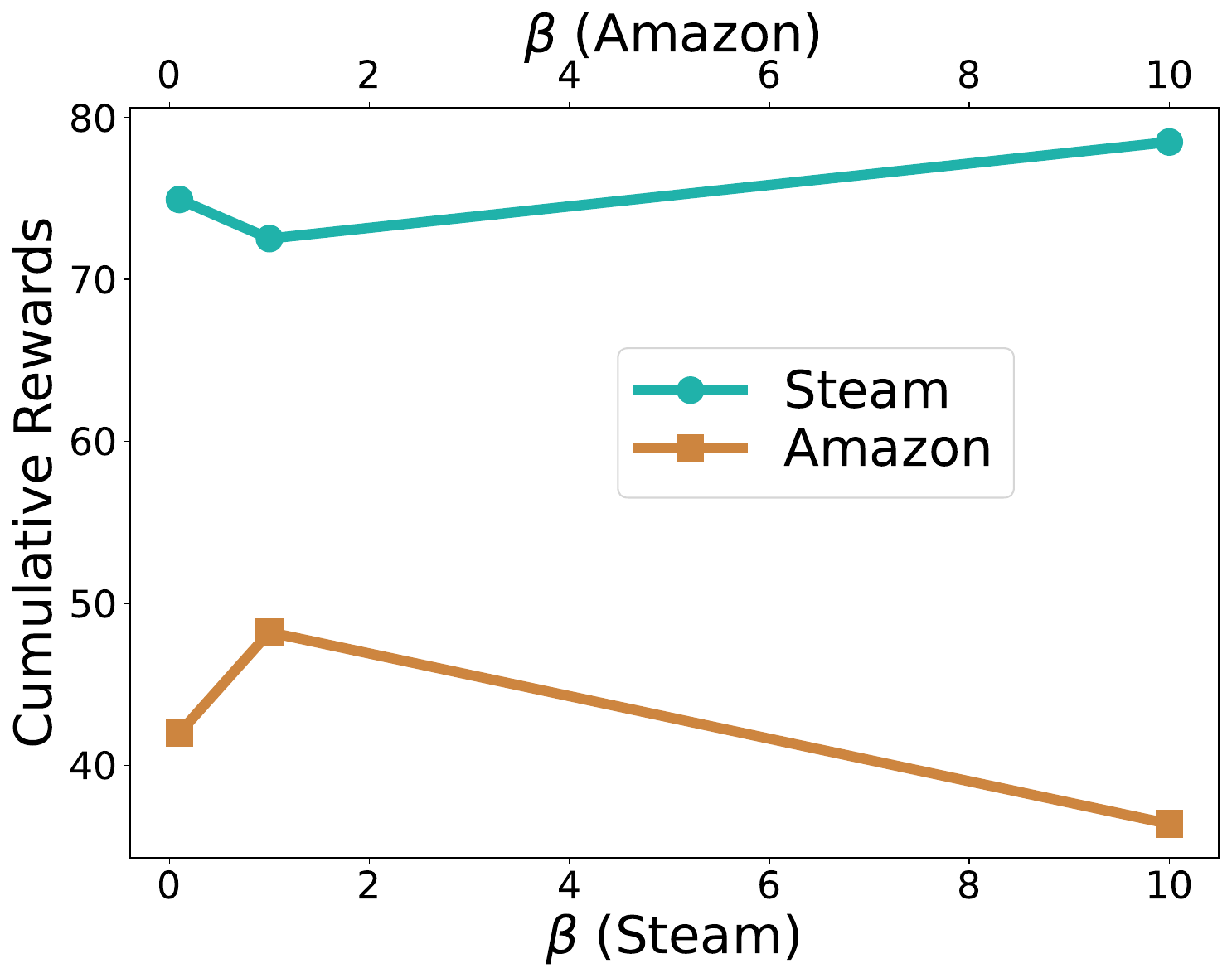}
    \end{minipage}
    \hfill
    \begin{minipage}[t]{0.485\columnwidth}
        \includegraphics[width=\columnwidth]{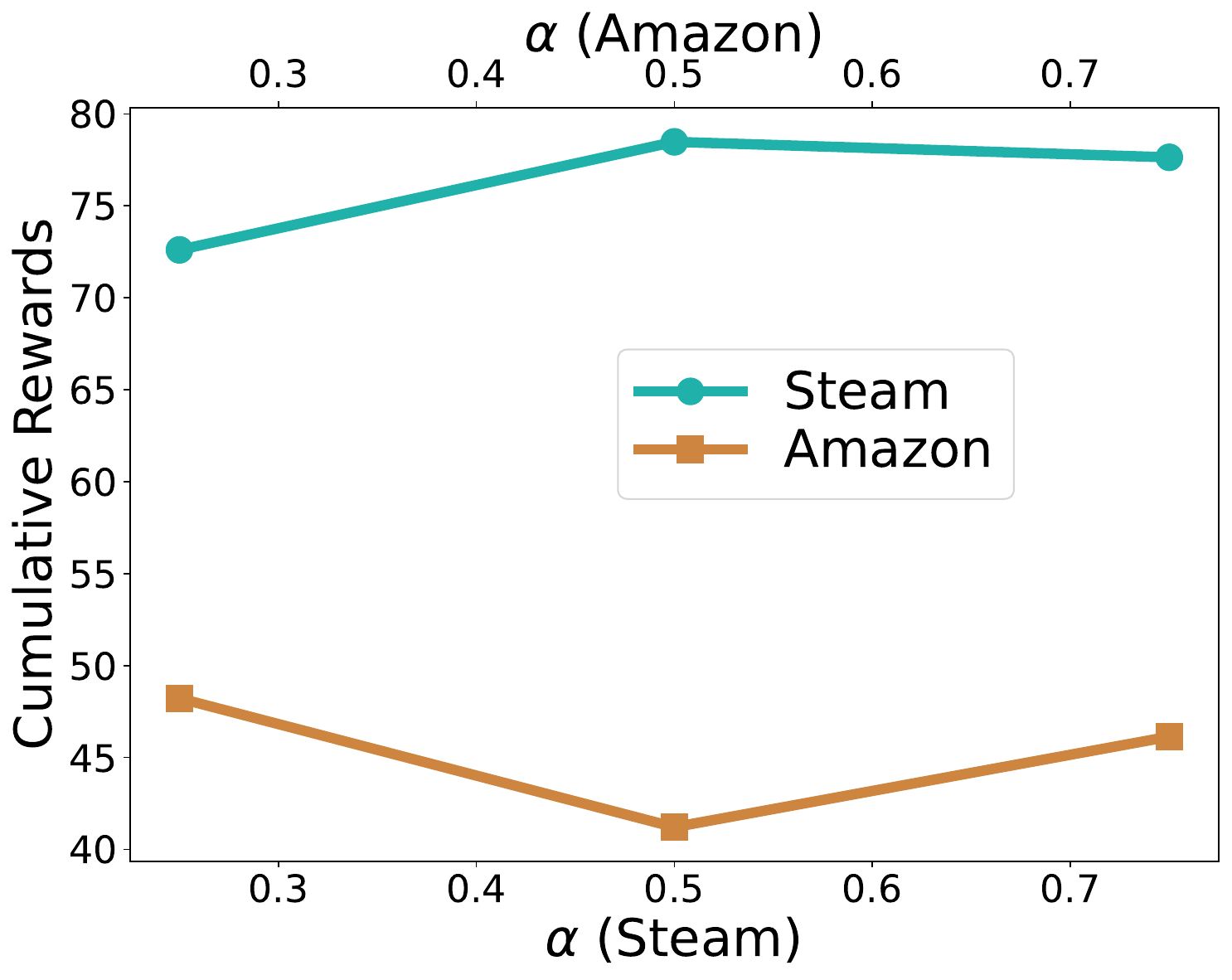}
    \end{minipage}
    
    \vspace{0.7em}  
    
    \begin{minipage}[t]{0.485\columnwidth}
        \includegraphics[width=\columnwidth]{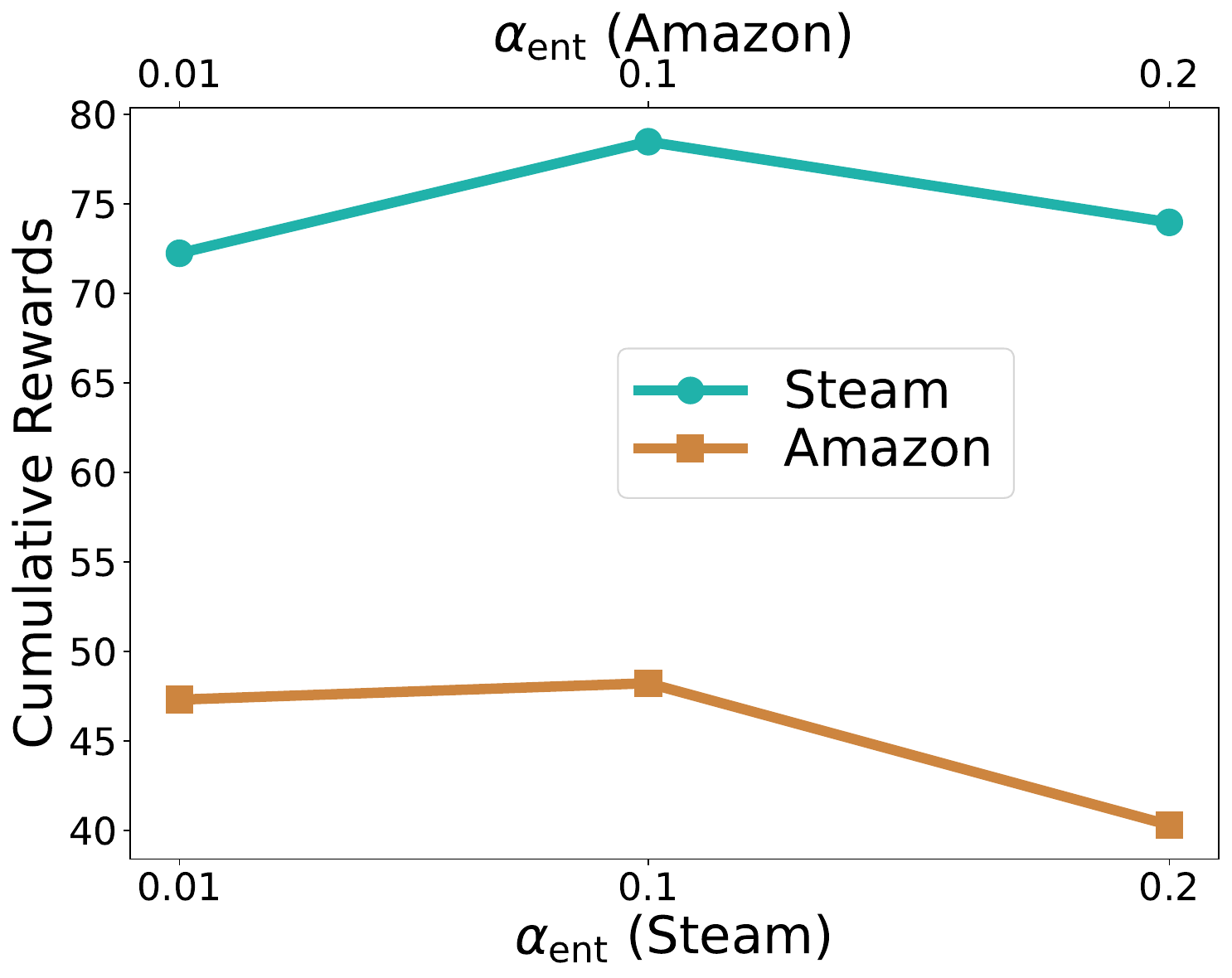}
    \end{minipage}
    \hfill
    \begin{minipage}[t]{0.485\columnwidth}
        \includegraphics[width=\columnwidth]{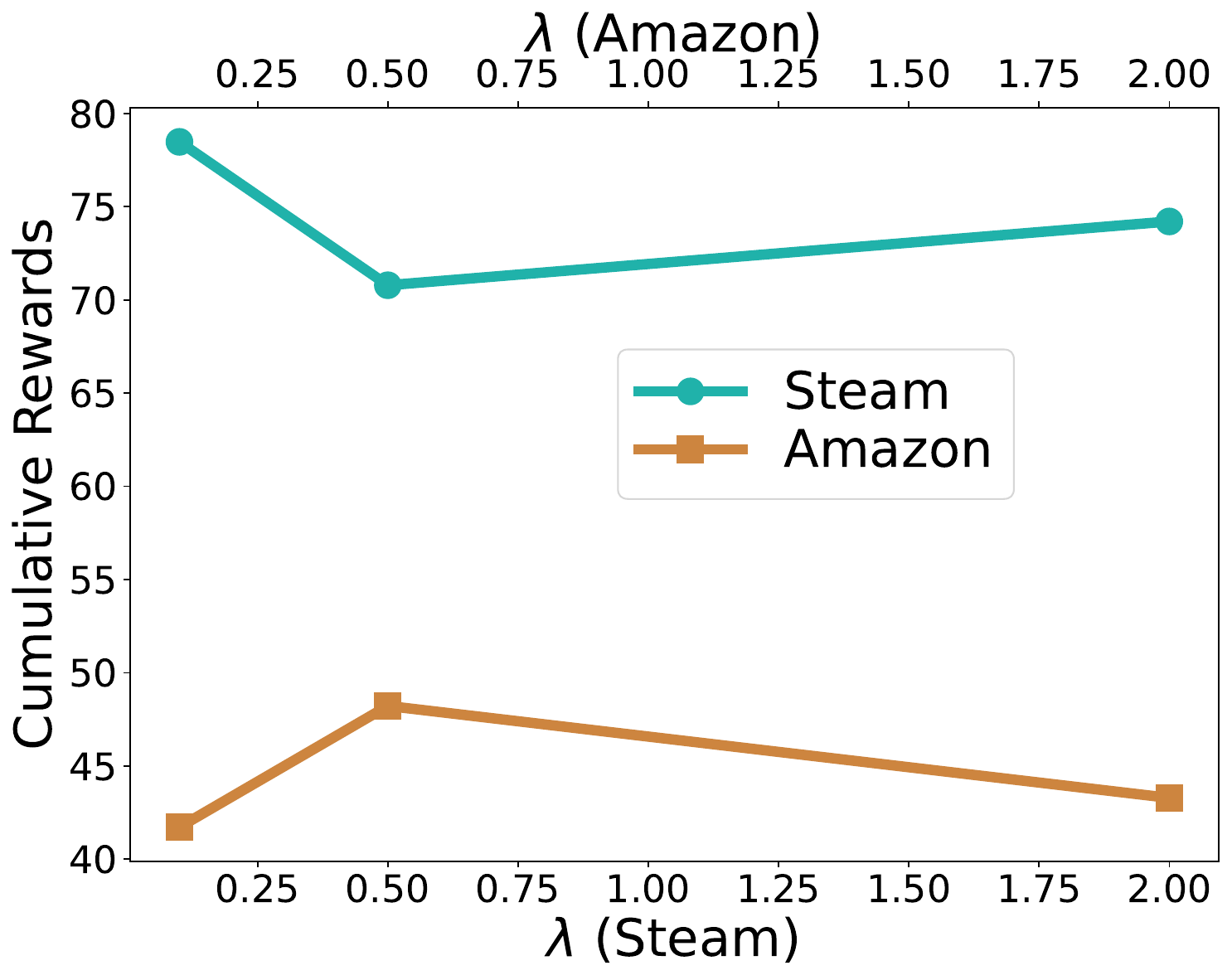}
    \end{minipage}
    
    \vspace{-0.5em}
    \caption{Hyperparameter sensitivity of $\beta$, $\alpha$, $\alpha_{\mathrm{ent}}$, and $\lambda_{\mathrm{imit}}$ on two datasets.}
    \label{fig:robust}
    \vspace{-1.2em}
\end{figure}

\vspace{-0.3em}
In this part, the robustness of IL-Rec against four critical hyperparameters is evaluated: the temperature $\beta$ in $w_{\mathrm{env}}$, the trade-off coefficient $\alpha$ in weight fusion, the entropy temperature $\alpha_{\mathrm{ent}}$ and the imitation loss weight $\lambda_{\mathrm{imit}}$. The corresponding results can be found in Fig.~\ref{fig:robust}.

\textbf{Effect of \(\beta\).} (Fig.~\ref{fig:robust}, top-left), the cumulative reward is seen to increase on Steam as \(\beta\) is raised, reaching its maximum at \(\beta=10\). On Amazon, the reward peaked at \(\beta=1\) before declining at higher values. This suggests that a larger temperature is beneficial for weighting environment advantage on Steam, while a moderate \(\beta\) is preferable for Amazon.

\textbf{Effect of \(\alpha\).} (Fig.~\ref{fig:robust}, top-right). On Steam, the highest reward was obtained at \(\alpha=0.5\), whereas on Amazon the best performance occurred at \(\alpha=0.25\). Values above or below these points led to reduced long-term returns, indicating the importance of balancing environment and IRL weights.

\textbf{Effect of \(\alpha_{\mathrm{ent}}\).} (Fig.~\ref{fig:robust}, bottom-left), the entropy temperature \(\alpha_{\mathrm{ent}}\) achieved optimal cumulative reward at 0.1 for both benchmarks. Lower or higher entropy weights resulted in modest performance drops, confirming that moderate exploration is critical to stable policy improvement.

\textbf{Effect of \(\lambda_{\mathrm{imit}}\).} The imitation loss coefficient \(\lambda_{\mathrm{imit}}\) was found to be most effective at dataset-specific values (Fig.~\ref{fig:robust}, bottom-right): \(\lambda_{\mathrm{imit}}=0.25\) on Steam and \(\lambda_{\mathrm{imit}}=0.5\) on Amazon. Larger weights yielded diminishing returns or slight performance degradation.

Overall, IL-Rec’s performance was shown to be relatively insensitive to variations in $\beta$, $\alpha$, $\alpha_{\mathrm{ent}}$, and $\lambda_{\mathrm{imit}}$. Although minor fluctuations in cumulative reward were observed, they remained confined to a narrow range, confirming the method’s robustness across reasonable hyperparameter settings.

\section{Conclusion and Future Work}
In this work, IL-Rec is proposed as a novel retrieval-enhanced hybrid RL framework that unifies imitation learning from LLM-generated demonstrations with independent RL policy refinement in a model-based offline recommendation paradigm. 
A discriminator-guided weighting mechanism is introduced, through which high-value transitions in demonstrations are selectively emphasized and the impact of sub-optimal ones is reduced, thereby enhancing recommendation policy learning.
Empirical results on Steam and Amazon benchmarks demonstrated that IL-Rec consistently outperforms both state-of-the-art offline RL and LLM-based baselines in cumulative reward and interaction length. 

Future directions include investigating more fine-grained aspects of LLM-generated demonstrations, such as their biases and coverage, to better understand their impact on policy learning. 
Extending IL-Rec to more diverse datasets across domains, real-world recommendation scenarios, and other sequential decision-making tasks is also part of the future work. 
In addition, adaptive strategies based on meta-learning or Bayesian optimization will be explored to automate hyperparameter tuning.

\section{Acknowledegments}
This work is supported by projects DE200101610, DE250100919, CE200100025 funded by Australian Research Council, and CSIRO’s Science Leader Project R-91559.

\bibliographystyle{IEEEtranS}
\bibliography{IEEEabrv,sample-base}

\appendix
\vspace{-0.3em}
\label{app:prompt}
\subsection{Demonstration of Prompts}
\vspace{-1em}
\begin{figure}[!h]
\centering
\includegraphics[trim=0cm 0cm 0cm 0cm, clip, width=1\linewidth]{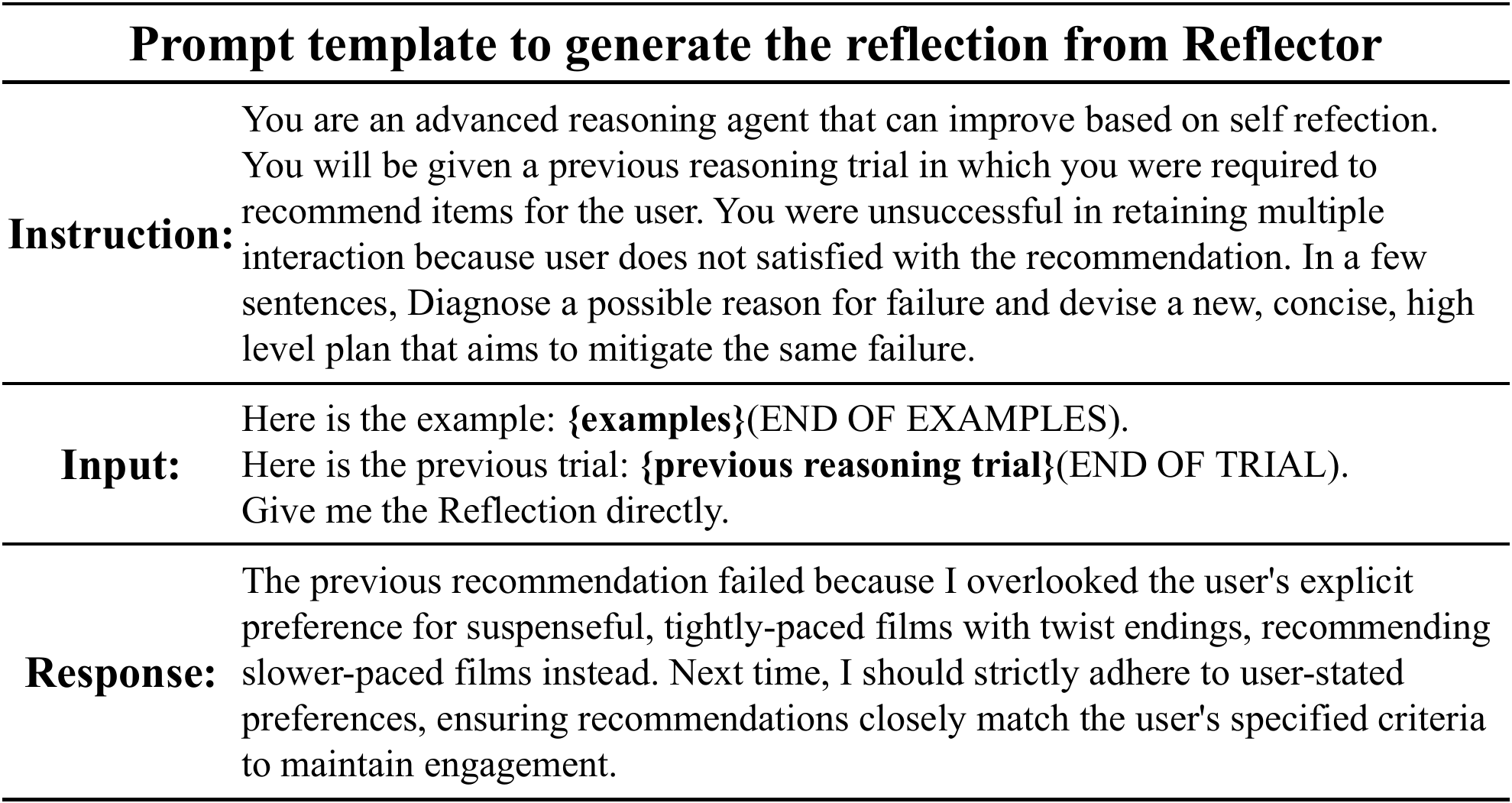}
\vspace{-1.8em}
\caption{Prompt to generate the reflection from Reflector as in BiLLP~\cite{shi2024billp}}
\label{fig:reflector}
\vspace{-1em}
\end{figure}

\begin{figure}[!h]
\centering
\includegraphics[trim=0cm 0cm 0cm 0cm, clip, width=1\linewidth]{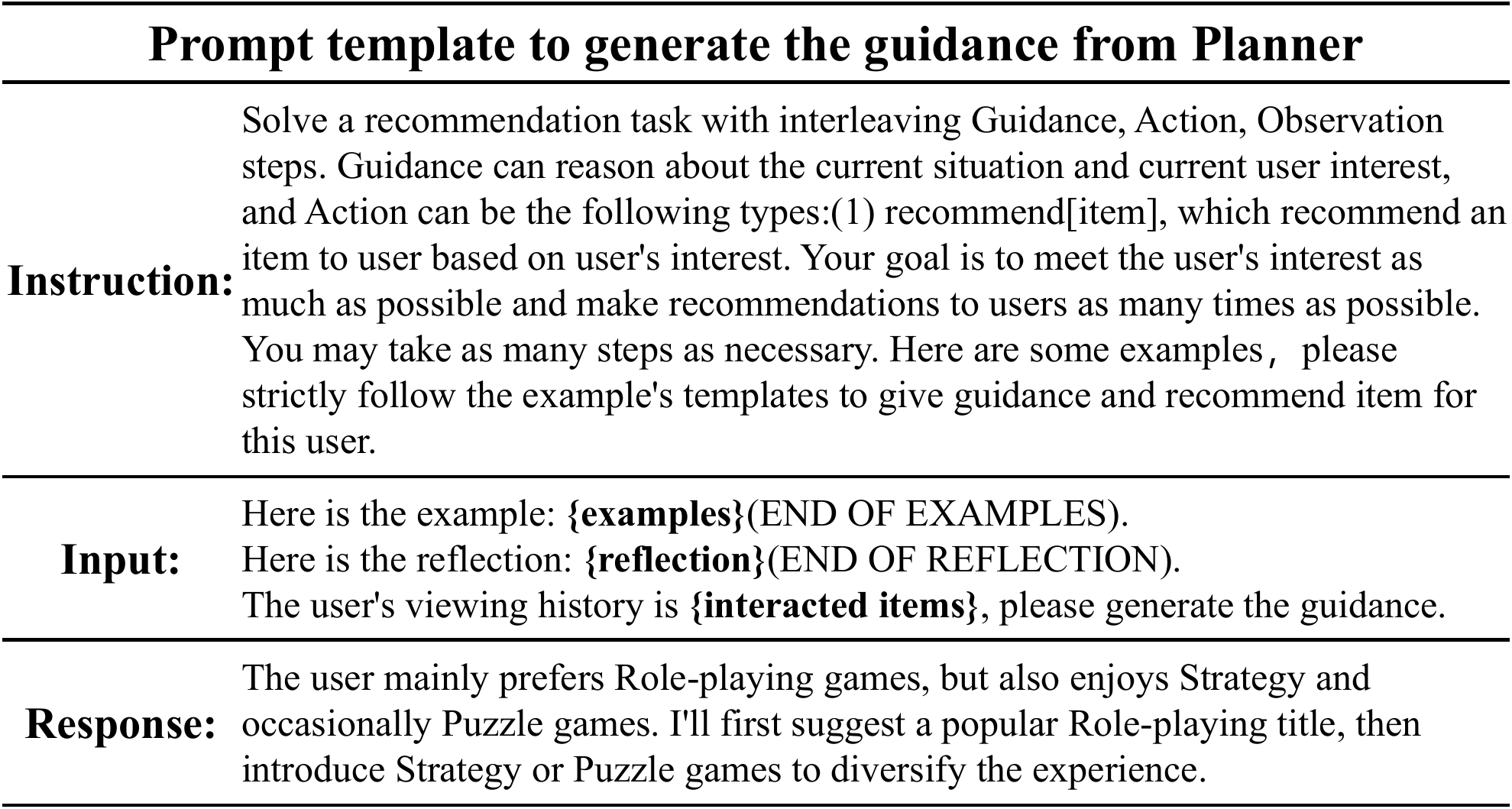}
\vspace{-1.8em}
\caption{Prompt to generate the guidance from Planner as in BiLLP~\cite{shi2024billp}}
\label{fig:planner}
\vspace{-1em}
\end{figure}

\begin{figure}[!h]
\centering
\includegraphics[trim=0cm 0cm 0cm 0cm, clip, width=1\linewidth]{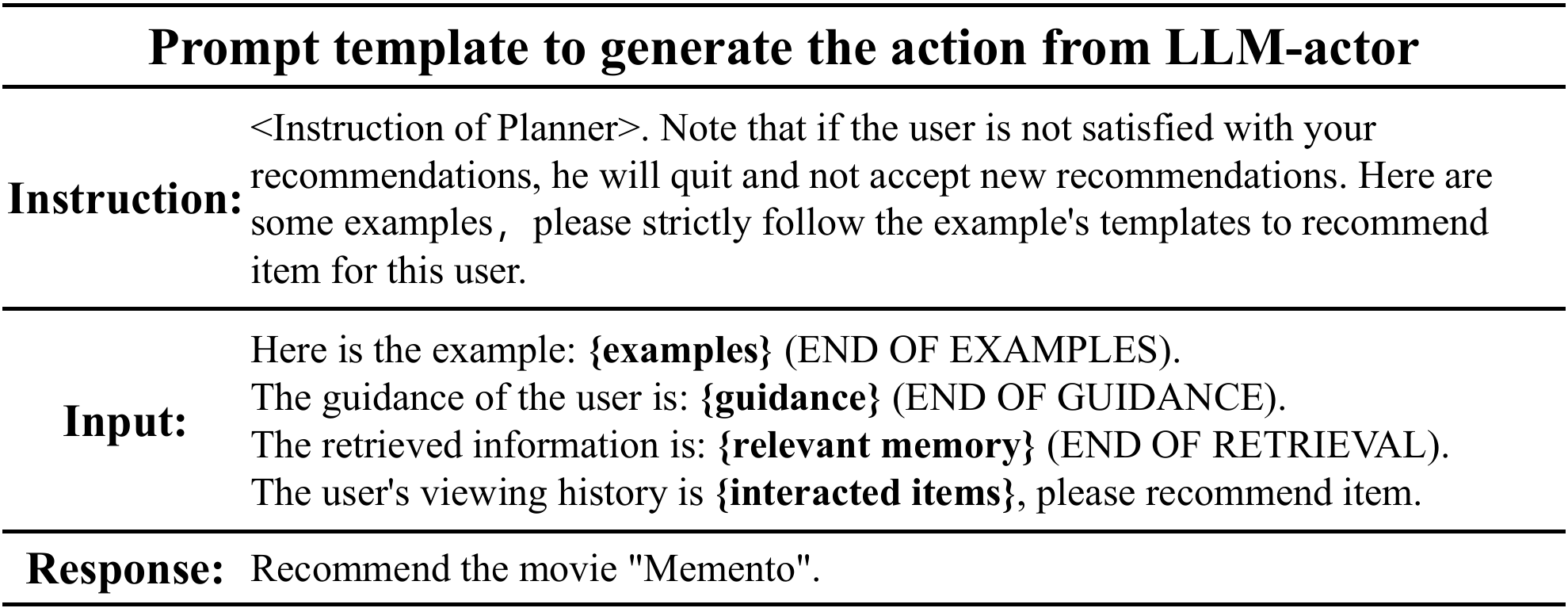}
\vspace{-1.8em}
\caption{Prompt to generate the action from LLM-actor as in BiLLP~\cite{shi2024billp}}
\label{fig:actor}
\vspace{-1em}
\end{figure}

\begin{figure}[!h]
\centering
\includegraphics[trim=0cm 0cm 0cm 0cm, clip, width=1\linewidth]{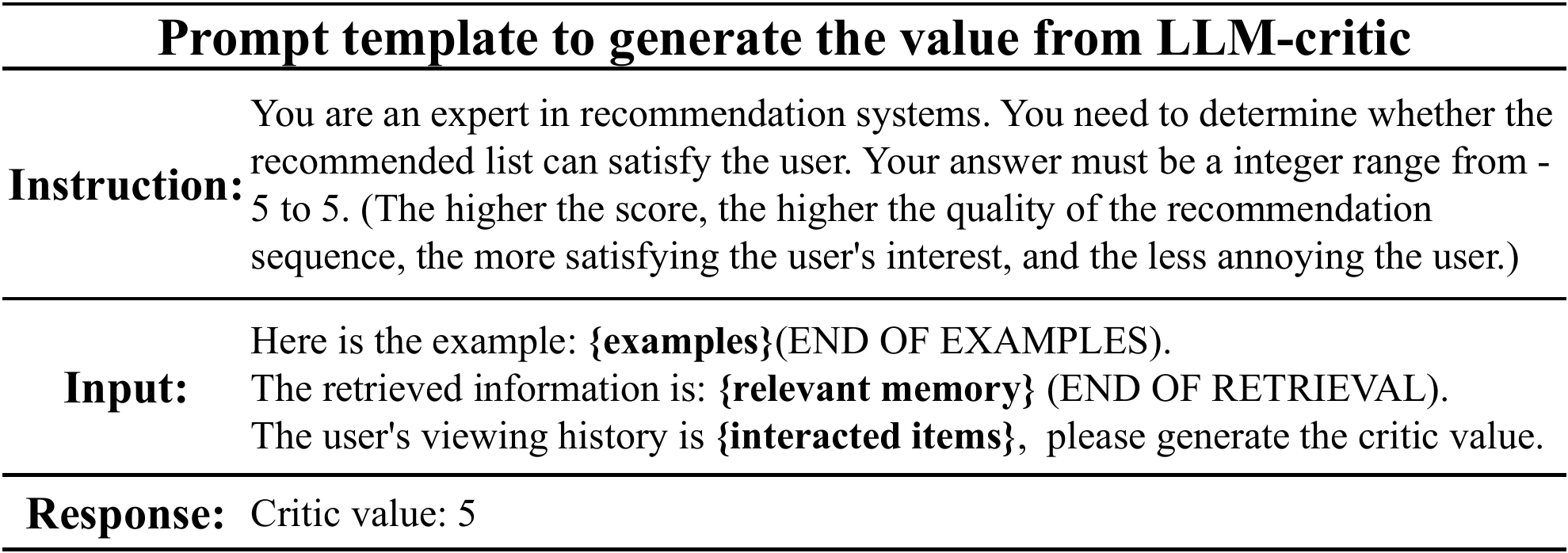}
\vspace{-1.8em}
\caption{Prompt to generate the value from LLM-critic as in BiLLP~\cite{shi2024billp}}
\label{fig:critic}
\vspace{-1em}
\end{figure}

\end{document}